\documentclass[conference]{IEEEtran}
\IEEEoverridecommandlockouts

\usepackage{amsmath,amssymb,amsfonts}
\usepackage{graphicx}
\usepackage{subcaption}
\usepackage{textcomp}
\usepackage[table]{xcolor}
\usepackage{booktabs}
\usepackage{array}
\usepackage{hyperref}
\usepackage{stfloats}
\usepackage{doi}
\usepackage{url}

\usepackage{orcidlink}
\usepackage[T1]{fontenc} 
\usepackage[scaled=0.86]{helvet}
\usepackage{booktabs}
\usepackage{soul}
\usepackage[noadjust]{cite}
\usepackage{xspace}
\usepackage{pbalance}

\newcommand{\PUCM}{\textsf{PM4Py-UCM\xspace}}
\newcommand{\PMPy}{\textsf{PM4Py\xspace}}
\newcommand{\jUCMNav}{\textsf{jUCMNav\xspace}}

\usepackage{tikz}

\definecolor{GreyUCM}{RGB}{71,85,105}
\definecolor{BlueUCM}{RGB}{29,78,216}
\definecolor{GreenUCM}{RGB}{15,118,110}
\definecolor{PurpleUCM}{RGB}{112,48,160}

\newcommand{\coloredcircle}[2]{%
  \tikz[baseline=(char.base)]{
    \node[shape=circle, draw=none, fill=#1, inner sep=1pt, minimum size=7pt] (char)
    {\textcolor{white}{\sffamily\bfseries\scriptsize #2}};
  }%
}

\newcommand{\One}[0]{\coloredcircle{GreyUCM}{1}}
\newcommand{\Two}[0]{\coloredcircle{BlueUCM}{2}}
\newcommand{\Three}[0]{\coloredcircle{GreyUCM}{3}}
\newcommand{\Four}[0]{\coloredcircle{GreenUCM}{4}}
\newcommand{\Five}[0]{\coloredcircle{GreenUCM}{5}}
\newcommand{\Six}[0]{\coloredcircle{PurpleUCM}{6}}
\newcommand{\Seven}[0]{\coloredcircle{PurpleUCM}{7}}
\newcommand{\Eight}[0]{\coloredcircle{PurpleUCM}{8}}

\usepackage{minted}
\begin{document}

\title{Towards Process Mining Use Case Map Models with \textsf{PM4Py-UCM}}

\author{%
\IEEEauthorblockN{Daniel Amyot\orcidlink{0000-0003-2414-1791}}
\IEEEauthorblockA{%
School of Electrical Engineering and Computer Science\\
University of Ottawa\\
Ottawa, Ontario, Canada\\
\texttt{damyot@uottawa.ca}}
}

\maketitle

\begin{abstract}
Given the increasing amount of data available in organizational systems, there is an opportunity for early requirements engineering (RE) activities to be better based on evidence than ever before. Process mining (PM) has been used for over two decades to discover and analyze as-is process models from event logs extracted from such data, with outputs often in the form of Petri Nets, directly-follows graphs, or BPMN models. This paper aims to make Use Case Map (UCM) models, from ITU-T's User Requirements Notation (URN), a first-class output of process discovery, so that mined behavior can be used in URN-based modeling, analysis, and management activities. This paper contributes and illustrates \PUCM, an open-source extension to the existing \PMPy{} Python library. This new tool contributes 1)~a UCM discovery pipeline, 2)~hierarchical decomposition strategies producing nested UCM models, 3)~configurable performer mappings for UCM and BPMN visualizations, and 4)~an exporter to a URN tool (\jUCMNav) that preserves the mined model under round-trip. Using public and synthetic event logs, the paper showcases how the same behavior is rendered under different performer abstractions and decomposition strategies, and discusses how PM can become a practical instrument for model-driven RE.

\textbf{\textit{Artifact---}}\url{https://github.com/ProcessMining-uOttawa/pm4py-ucm}
\end{abstract}


\section{Introduction}
\label{sec:intro}

\textit{Process mining} (PM), which enables the automated discovery and analysis of process models from event logs, has matured into a discipline that delivers data-driven evidence of how organizations actually execute their processes~\cite{vanderAalst2016}. 
In particular, PM is known to support the discovery of process models in various notations, some limited to sequencing and alternatives (e.g., Directly-Follows Graphs -- DFGs) and others also support concurrency (e.g., Petri nets -- PNs~\cite{PetriNets}, and OMG's Business Process Model and Notation -- BPMN~\cite{BPMN}). Model elements are often annotated with frequency, time, or cost annotations, hence enabling performance analysis and the elicitation of improvement requirements. Discovered models, often labeled \textit{as-is models}, can also be compared to expected or prescribed models during conformance analysis, to assess differences and infer alignment requirements. A few methodologies have emerged over the years to take advantage of PM in a process improvement context, with $PM^2$ being the best-known one~\cite{PM2}.

Dozens of PM tools now exist on the market, most being cloud-based and commercial~\cite{Kesici2022PMToolsSLR}. Among the open-source PM solutions available, one stands out as increasingly popular: \PMPy~\cite{Berti2023-pm4py}. This Python library is used by practitioners and researcher alike due to its free access, the availability of its code, its extensibility, and the fact that it can run on local computers (an important capability for organizations that want to avoid sharing event logs on the cloud, often for privacy reasons). 

From a requirements engineering angle, PM can be seen as a data-based elicitation technique~\cite{Lim2021-Data-Driven-RE-SLR} that aligns well with crowd-based requirements engineering when the source event logs capture user interactions with systems. D\k{a}browski et al.~\cite{Dabrowski2017} exploited that relationship to propose a method combining requirements engineering, process mining, and crowdsourcing to discover the underlying processes of crowds from event logs, and detect misalignments with expected system behavior. More recently, Ghasemi and Amyot~\cite{Ghasemi2018-RE-PM,Ghasemi2025-GoPM} explored the use of goal models as a filtering mechanism for preserving traces in an event log that satisfy a related collection of measurable goals. Bouhidel et al.~\cite{Bouhidel2026-RE-for-PM} suggested a requirements-driven methodology that emphasizes specificity and iterative refinement when discovering models and presenting results, while calling for the customization of PM tools and analytical assets. 

This paper answers this call by proposing a tool-oriented bridge between the process mining ecosystem and the requirements engineering ecosystem around the Use Case Map (UCM) notation, part of the User Requirements Notation (URN) standard~\cite{ITU2018-URN}. The UCM notation, originally proposed by Buhr in the 1990's~\cite{Buh98}, is combined with the Goal-oriented Requirement Notation (GRL) in URN, enabling the automated analysis of goal-process alignment, something that PNs and BPMN do not support on their own, out of the box. URN's UCM notation also supports the binding of process activities (called \textit{responsibilities}, akin to BPMN activities) to two-dimensional hierarchical structures of components (e.g., systems, system modules, users, and other types of \textit{performers}), going beyond the stricter pool representation in BPMN.

In order to close the existing gap between process mining and the UCM/URN ecosystem, this paper extends \PMPy{} with support for the UCM notation, leading to the \PUCM{} tool. More precisely, the contributions of this paper are:
\begin{itemize}
  \item \textbf{C1.} An event-log-to-UCM discovery pipeline with a documented mapping from \PMPy's intermediate structures to UCM constructs. A Web interface to a deployed version of the pipeline (with sample event logs) is available online\footnote{Web interface demo: \url{https://pm4py-ucm.streamlit.app/}}.
  \item \textbf{C2.} Configurable hierarchical decomposition strategies producing nested UCM models, also with support for BPMN (visualization only).
  \item \textbf{C3.} Performer-aware bindings (i.e., responsibilities connected to their performing components), with visualization for UCM and BPMN models that can be exported to PNG files.
  \item \textbf{C4.} A bi-directional importer/exporter that supports the \jUCMNav{} file format (\texttt{.jucm}, XML-based).
\end{itemize}

The rest of this paper is organized as follows. Section~\ref{sec:background} provides background information on the UCM notation and on process mining with \PMPy. Section~\ref{sec:architecture} then introduces \PUCM's architecture and discovery pipeline while Section~\ref{sec:capabilities} explores and illustrates RE-related capabilities. Sections~\ref{sec:related} and~\ref{sec:limitations} discuss related work and limitations, respectively, while Section~\ref{sec:conclusion} concludes.

\section{Background}
\label{sec:background}

\subsection{Use Case Map Notation and jUCMNav}
\label{sec:UCM}

Use Case Maps, standardized by the International Telecommunication Union (ITU-T)~\cite{ITU2018-URN}, are a requirements notation used for modeling and analyzing processes and scenarios. As summarized in Table~\ref{tab:notations}, the main UCM concepts share many similarities with those of BPMN~\cite{BPMN}, especially around start/end points, activities (responsibilities in UCM), sequencing, choice, concurrency, decomposition, and performers (components in UCM, pools in BPMN). This common subset is often the one targeted as output by process mining algorithms. BPMN also offers concepts not supported by UCMs, including message flows, object flows, and many more types of start/end events, which are often used when BPMN models specify executable automations. Meanwhile, UCMs offer an executable semantics based on a simple data model that supports scenario-based testing and visualization as part of the language. Being part of URN, UCMs are also integrated with and traceable to GRL goal models. These concepts unique to UCMs are more useful for requirements engineering activities than the ones unique to BPMN; however, they are not targeted by the current version of \PUCM.

\begin{table}[h]
    \caption{Simple Comparison of Process Notations}
    \centering
    \scriptsize
    \rowcolors{2}{white}{blue!15}
    \label{tab:notations}
    \begin{tabular}{lllll}
        \toprule
        \textbf{Concept} & \textbf{BPMN} & \textbf{Petri Nets} & \textbf{DFG} & \textbf{UCM} \\
        \midrule
        \textbf{Star/End}        & Yes & Yes & Yes       & Yes \\
        \textbf{Activity}        & Activity/Task & Place & Activity       & Responsibility \\
        \textbf{Sequence}        & Yes               & Yes   & Yes            & Yes            \\
        \textbf{Choice}          & Yes               & Yes   & Yes    & Yes \\
        \textbf{Concurrency}     & Yes               & Yes   &                & Yes            \\
        \textbf{Decomposition}   & Call Activity     & Subnet   &                & Stub            \\
        \textbf{Performer}            & Pool/Lane     &       &                & Component      \\
        \textbf{Artifact}        & Object/Data Store   & Col. token      &                & Object               \\
        \textbf{Message flows}   & Yes               &       &                &                \\
        \textbf{Links to Goals}  &                   &       &                & Yes            \\
        \textbf{Data Model}      &                   &       &                & Yes            \\
        \textbf{Test Definitions}&                   &       &                & Yes            \\
        \textbf{Standardized}    & OMG               & ISO   &                & ITU-T    \\
        \bottomrule
    \end{tabular}
\end{table}

Two decades of experience in using URN for different requirements engineering activities, including goal/process modeling and alignment, integration with requirements management systems, regulatory compliance and intelligence, process adaptation and improvement, value co-creation, and advanced extensions for feature modeling, aspect-oriented modeling, and model slicing, have recently been summarized~\cite{Amyot2022-EMISAJ} and could hence be made available to requirements engineers interested in PM-based process discovery. This is part of the added-value offered by the support of UCM models in a PM context.

Many of these functionalities are currently supported by \jUCMNav~\cite{RKA06}, an Eclipse-based environment that is the de facto standard for URN modeling and analysis. jUCMNav uses an XML-based file format for URN models, which is a serialization of the objects instantiating the tool's metamodel. This open-source tool is freely available online~\footnote{\url{https://github.com/JUCMNAV/jUCMNavPlus/}}.

\subsection{Process Mining and PM4Py}
Process mining uses event logs as input, and uses a mining algorithm that produces a process model (often with annotations) as output. An event log is generally produced by transforming event-related information found in information systems. A log is a list of events where each event must minimally contain information about its \textit{case} identifier (customer, patient visit, sales transaction, etc.) capturing the process instance, together with the \textit{activity} name and its \textit{timestamp}. Other optional attributes can be used for enabling filtering and the computation of annotations (e.g., average activity costs) on the resulting model. Of particular here are performer-oriented attributes such as \textit{roles} (e.g., client or claims officer) and particular \textit{resources} (e.g., Nancy or Gary) playing such roles. Roles and resources are candidate attributes for representation as components in UCM models. Event logs are often stored as Comma-Separated Value (CSV) files or as standard XES files~\cite{XES2024}, where XES uses XML to provide clearer semantic definitions of the attributes.

Different mining algorithms exist, which offer different trade-offs~\cite{vanderAalst2016}. Some are quick but do support concurrency: $\alpha$-miner is precise but does not provide much abstraction and leads to complex models, whereas the heuristic-miner and fuzzy-miner algorithms provides some abstraction and better understandability. Others are slower but support  concurrency detection (leading to simpler models, albeit less precise), including the inductive-miner and the split-miner. The \textit{inductive-miner} algorithm~\cite{LFvdA13a,LFvdA13b} is of particular interest here as it generates well-nested process trees, amenable to representation with BPMN and UCM, and with a simple path towards support for hierarchical process decomposition. This algorithm first builds a DFG and then identifies a \textit{cut} (a structural split) in that graph that matches basic workflow patterns, namely:
\begin{itemize}
    \item Sequence (\(\rightarrow \)): Activity \textsf{A} must finish before activity \textsf{B} starts.
    \item Exclusive Choice (\(\times \)): Either activity \textsf{A} or activity \textsf{B} happens, but not both.
    \item Inclusive Choice (\(\lor \)): Activity \textsf{A} or activity \textsf{B} happens, or both happen.
    \item Parallel (\(\land \)) and Interleaving (o): Activities \textsf{A} and \textsf{B} happen concurrently, or in any order.
    \item Loop (\(\circlearrowleft \)): An activity can be repeated multiple times.
    \item Silent (\(\tau \)): An invisible activity, often used to capture an internal routing mechanism or skip logic.
\end{itemize}
Note that the above activities can also be sub-processes. Then, the algorithm splits the log at that cut point, and recursively applies the same split-based decomposition on each sub-log.

Although dozens of PM solutions exist nowadays~\cite{Kesici2022PMToolsSLR}, \PMPy{} is relevant here as this is an open-source and extensible library that supports several mining algorithms (including the inductive-miner\footnote{See documentation of \href{https://pm4py-source.readthedocs.io/en/stable/pm4py.html?highlight=inductive\#pm4py.discovery.discover\_process\_tree\_inductive}{\texttt{discover\_process\_tree\_inductive}}.}), as well as existing transformations to BPMN models~\cite{Berti2023-pm4py}. However, none of these tools, including \PMPy, supported the UCM notation before this paper.

\section{\PUCM: Discovery Pipeline}
\label{sec:architecture}
The main architectural strategy behind \PUCM{} is to reuse \PMPy{} for discovering a process tree, and then provide a transformation to a UCM object model (implemented as Python objects) that can be used for UCM visualization (akin to the existing BPMN ones) and model export to jUCMNav (\texttt{.jucm} file). The \textit{UCM object model}, available online\footnote{UCM object model available on GitHub in \href{https://github.com/ProcessMining-uOttawa/pm4py-ucm/blob/main/docs/ucm_class_diagram.png}{PNG} and \href{https://raw.githubusercontent.com/ProcessMining-uOttawa/pm4py-ucm/b07e44b792b39f51ef7c2fe73505de258d01fae5/docs/ucm_class_diagram.svg}{SVG}.}, describes the various types of UCM path elements and their attributes, as well as how they are grouped in UCM maps, how stubs are decomposed into sub-maps, and how responsibilities are bound to components.

\begin{figure*}[t]
    \centering
    \includegraphics[width=1\textwidth]{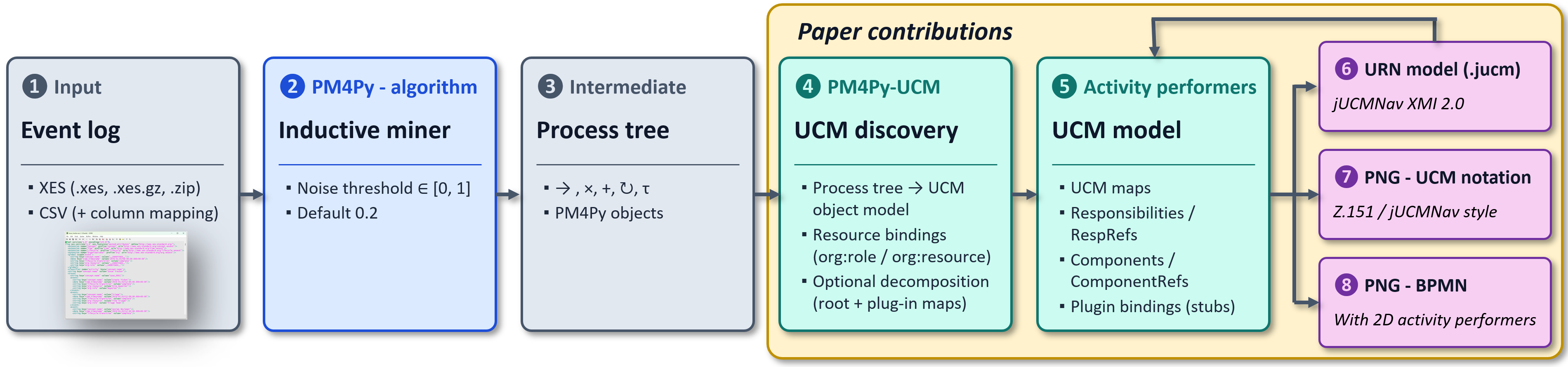}
    \caption{The \textsf{PM4Py-UCM} discovery pipeline.}
    \label{fig:pipeline}
\end{figure*}

Figure~\ref{fig:pipeline} gives an overview of the discovery pipeline, which is this paper's first contribution (\textbf{C1} in the introduction). Steps \One, \Two, and \Three{} are performed with existing \PMPy{} features, steps \Four{} and \Five{} are where the main contributions of this paper reside, whereas artefacts \Six, \Seven, and \Eight{} show the different outputs generated. From an event log in CSV or XES format \One, \PMPy's inductive-miner algorithm is called \Two{} to produce a process tree \Three. The algorithm has a threshold parameter that can be set, to enable the mining of concurrency to be more or less tolerant to noise in the event log. That algorithm is used as is in the pipeline.

In step~\Four, the UCM discovery step converts the mined process tree to a UCM model that instantiates, using the mapping in Table~\ref{tab:mapping}, the object model highlighted earlier. At this point, there is neither process decomposition nor performer binding done; this is in a way akin to current \PMPy{} capabilities that generate BPMN models from process trees~\cite{Berti2023-pm4py}. 

\begin{table}[hb]
    \centering
    \caption{Mapping from Process Tree Constructs to UCM Constructs}
    \label{tab:mapping}
    \scriptsize
    \rowcolors{2}{white}{blue!15}
    \begin{tabular}{p{0.20\columnwidth} p{0.70\columnwidth}}
        \toprule
        \textbf{Process Tree} & \textbf{UCM Construct} \\
        \midrule
        Activity \textsf{A} &
        \texttt{RespRef} referencing a \texttt{Responsibility} \textsf{A} \\
        $\rightarrow$ &
        Children chained with \texttt{EmptyPoint} connectors (some being collapsed by a model simplifier) \\
        $\times$ and $\lor$ &
        \texttt{OrFork} $\rightarrow$ branches $\rightarrow$ \texttt{OrJoin} \\
        $\land$ and o &
        \texttt{AndFork} $\rightarrow$ branches $\rightarrow$ \texttt{AndJoin} \\
        {$\circlearrowleft$} &
        \texttt{OrJoin} $\rightarrow$ \texttt{do} $\rightarrow$ \texttt{OrFork} with \texttt{[redo]} back-edge and \texttt{[exit]} forward edge \\
        $\tau$ &
        Direct \texttt{NodeConnection} with no responsibility \\
        \bottomrule
    \end{tabular}
\end{table}

The second contribution of this paper (\textbf{C2}), also found in step \Four, focuses on the hierarchical decomposition of process models. The decomposition algorithm starts with a process tree-level planning step (i.e., to pick which subtrees become UCM plug-in maps for parent stubs) followed by a conversion-time intercept (i.e., the cut handler diverts those subtrees into their own maps as they are converted). The planning decisions are influenced by two categories of parameters: 
\begin{enumerate}
    \item \textbf{Workflow patterns of focus} (top four rows of Table~\ref{tab:decomposition}): these Boolean options enable users to decide which sequencing operators should be considered as potential cut points for decomposition. The first one is for the root map only (to handle long sequences of activities) whereas the others apply to all levels of decomposition. 
    \item \textbf{Size-based requirements} (bottom three rows of Table~\ref{tab:decomposition}): these numerical parameters enable users to set minimum and maximum numbers of elements (activities/stubs) per map, to better handle complexity. They can influence the decomposition depth.
\end{enumerate}

\begin{table}[t]
    \centering
    \caption{Process Decomposition Parameters}
    \label{tab:decomposition}
    \scriptsize
    \rowcolors{2}{white}{blue!15}
    \begin{tabular}{p{0.27\linewidth}p{0.07\linewidth}p{0.51\linewidth}}
        \toprule
        \textbf{Parameter} & \textbf{Default} & \textbf{Meaning} \\
        \midrule
        \textsf{on\_root\_sequence} & \texttt{True} &
        Each child of a top-level $\rightarrow$ becomes a plug-in. Root map reads as a chain of phase stubs. \\
        
        \textsf{on\_parallel} & \texttt{True} &
        Each $\land$ branch becomes a plug-in. AND-fork/join vertical-expansion cost is replaced by a single stub per branch. \\
        
        \textsf{on\_alternative} & \texttt{True} &
        Each $\times$/$\lor$ branch becomes a plug-in. OR-fork/join stays on the parent map; alternative bodies move into per-branch plug-ins. \\
        
        \textsf{on\_loop} & \texttt{True} &
        Each $\circlearrowleft$ operator's expansion becomes a plug-in. Parent map reads as forward flow with one stub for the iteration. \\
        
        \textsf{max\_leaves\_per\_map} & \texttt{20} &
        Cap: over-sized maps recursively force-cut the largest operator-subtree until the cap is met. Can override the above rules. \\
        
        \textsf{min\_leaves\_to\_decompose} & \texttt{4} &
        Floor: subtrees smaller than this stay inlined regardless of rules. \\
        
        \textsf{balance\_ratio} & \texttt{0.2} &
        Sibling share threshold under $\rightarrow$ and $\land$. A child needs at least this fraction of the parent's leaves to be pulled out independently. \\
        \bottomrule
    \end{tabular}
\end{table}

As a convenience, several predefined decomposition modes (Table~\ref{tab:decompositionModes}) are offered to users:
\begin{table}[ht]
    \centering
    \caption{Process Decomposition User Modes}
    \label{tab:decompositionModes}
    \scriptsize
    \rowcolors{2}{white}{blue!15}
    \begin{tabular}{p{0.12\linewidth}p{0.78\linewidth}}
        \toprule
        \textbf{Mode} & \textbf{Definition} \\
        \midrule
        \textsf{None / "off"} &
        No decomposition. \\
        
        \textsf{"auto"} &
        All workflow-related rules set to \texttt{True}, with \textsf{max\_leaves\_per\_map=20} and \textsf{min\_leaves\_to\_decompose=4}. \\
            
        \textsf{"aggressive"} &
        Same as \textsf{"auto"} but with \textsf{max\_leaves\_per\_map=10}. \\
        
        \textsf{dict} &
        A dictionary entry that provides specific values for any combination of the seven parameters in Table~\ref{tab:decomposition}. \\
        \bottomrule
    \end{tabular}
\end{table}

Hierarchical UCM models can now be visualized and exported in PNG format \Seven, and exported valid \texttt{.jucm} files readable by \jUCMNav{} \Eight. At this point, a simple script that uses \PUCM{} to mine a log in XES, decompose it aggressively, and save it both as a diagram and as a \jUCMNav{} file is as follows:
\begin{minted}[style=solarized-dark, bgcolor=black, fontsize=\footnotesize]{python}
import pm4py
import pm4py_ucm

log = pm4py.read_xes("log.xes")
ucm = pm4py_ucm.discover_ucm_inductive(log, 
         decomposition="aggressive")

pm4py_ucm.save_vis_ucm(ucm, "diagram_ucm.png")
pm4py_ucm.write_ucm(ucm, "log.jucm")
\end{minted}

Note that \PMPy{} also enables users to create process trees and UCM models programmatically (i.e., through Python method calls), which is useful for testing purposes. In addition, jUCMNav files can be modified and re-imported again as Python UCM model (see arrow from \Six{} to \Five{} in Figure~\ref{fig:pipeline}), ensuring support for round-trip engineering. This essentially defines contribution \textbf{C4}.

Step \Five{} is concerned with contribution \textbf{C3}. At this point, UCM components are still not identified in the process model. The strategy here is to harvest roles and resources from an event log, assess what activities are related to them, and inject a simplified mapping (each activity being bound to one component, or none) into the UCM model.

\begin{minted}[style=solarized-dark, bgcolor=black, fontsize=\footnotesize]{python}
# One-shot: mine and bind in a single call.
ucm = pm4py_ucm.discover_ucm_inductive(log, 
         parameters={"resource_attribute": 
              ["org:role", "org:resource"],
})
pm4py_ucm.write_ucm(ucm, "log.jucm")
\end{minted}

This performer-binding information can also be provided manually, through Python method calls; this capability is again useful for testing, but also for considering other sources of mapping information outside the log. The performer binding can use the role, resource, or any other attribute from the log as a source of UCM components. 

As \PUCM{} is currently limited to one component per activity, and as the same activity in a log can be performed by many roles or resources, several options are offered when extracting performer bindings from a log (see Table~\ref{tab:performer-strategies}), which uses an example where activity \emph{Act} with 10 events, performed by Alice (5$\times$), Bob (3$\times$), Carol (2$\times$, first in log order).

\begin{table}[h]
  \centering
  \small
  \caption{Activity$\to$Performer Aggregation Strategies}
  \label{tab:performer-strategies}
  \scriptsize
  \rowcolors{2}{white}{blue!15}
  \begin{tabular}{p{0.1\linewidth}p{0.57\linewidth}p{0.16\linewidth}}
        \toprule
    \textbf{Strategy} & \textbf{Definition} & \textbf{Output} \\
    \midrule
    \textsf{mode} \emph{(default)} &
      Most-frequent performer, with ties broken lexicographically.
      Activity left unbound if modal share $<$ \textsf{min\_support}. & \texttt{Alice}\,\textsuperscript{$\star$} \\
    \textsf{first} &
      Performer of the first event for the activity, in log order. & \texttt{Carol} \\
    \textsf{unbound} &
      Performer iff exactly one distinct value is observed;
      activity is omitted otherwise. & \emph{(omitted)} \\
    \textsf{all} &
      Composite \texttt{"p\textsubscript{1}+p\textsubscript{2}+\ldots"}
      listing every distinct performer. & \texttt{Alice+ Bob+Carol} \\
    \bottomrule
  \end{tabular}
  \footnotesize \textsuperscript{$\star$}\,Omitted if \texttt{min\_support} $>$ 0.5 (modal share is 5/10).
\end{table}

Once injected into the UCM object model, this performer binding information is used to create UCM components as well as bindings between activities and their containing components. 

Going back to model generation, the components are also visualized in the PNG bitmaps \Seven{} and exported to the \texttt{.jucm} files \Six. Component colors are also selected based on the hashed value of the component name, so that the component remains colored the same way across generated models. Decomposed processes are stacked vertically with meaningful name inferred for stubs and their plug-ins. 

As the UCM object model is in many ways similar to BPMN's in its support for activity sequences, choices, concurrency, decomposition, and performers (as seen in Table~\ref{tab:notations}), \PUCM{} also offers a BPMN-based hybrid PNG visualization \Eight, where the BPMN syntax is used to describe processes whose activities are bound and shown visually as UCM components instead of as pools and lanes. Activities are shown in yellow, decomposed activities in pink with a \textbf{+} annotation, gateways are in green, and the same coloring approach used for UCM models is also used for BPMN models. However, no real BPMN model (in XMI) is generated for now.

\section{Illustration of RE-Relevant Capabilities}
\label{sec:capabilities}
Two event logs will be used to illustrate the main capabilities discussed in the previous section, one about an issue tracking process (synthetic log with 100,008 events, 11,284 cases, and 9 activities), and the other about a claims payment process (subset of an anonymized real log with 78,126 events, 5,600 cases, and 25 activities). Both are available online on the \PUCM{} GitHub repository\footnote{\url{https://github.com/ProcessMining-uOttawa/pm4py-ucm/}} and are used in two provided demo artifacts: a tutorial in \textit{Jupyter Notebook}, and a Web-based interface built in Python on top of \textit{Streamlit}~\cite{Streamlit2025}. The main RE scenario here pertains to the data-driven elicitation of a process model, and especially "who does what, and when".

\begin{figure*}[t]
    \centering

    \begin{subfigure}{\textwidth}
        \centering
        \includegraphics[width=\textwidth]{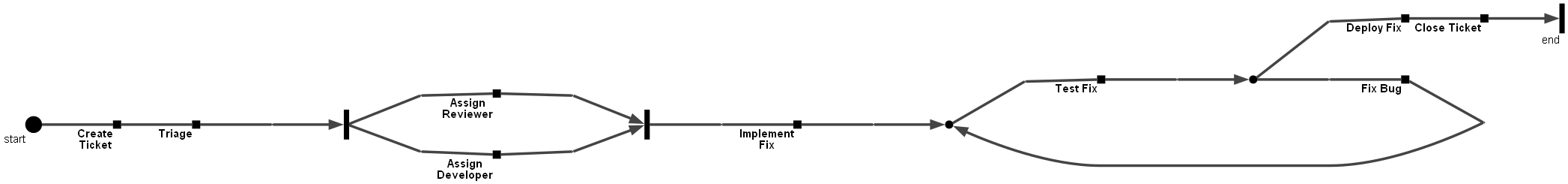}
        \caption{ UCM view, no component}
        \label{fig:initial-ucm-nocomp}
    \end{subfigure}

    \vspace{0.5em}

    \begin{subfigure}{\textwidth}
        \centering
        \includegraphics[width=\textwidth]{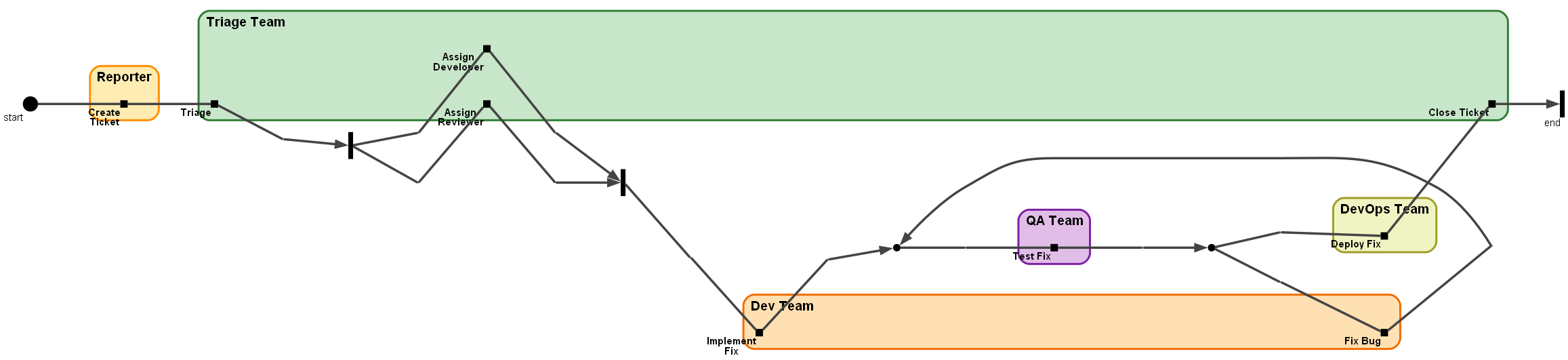}
        \caption{UCM view, with roles as component}
        \label{fig:initial-ucm-roles}
    \end{subfigure}

    \vspace{0.5em}

    \begin{subfigure}{\textwidth}
        \centering
        \includegraphics[width=\textwidth]{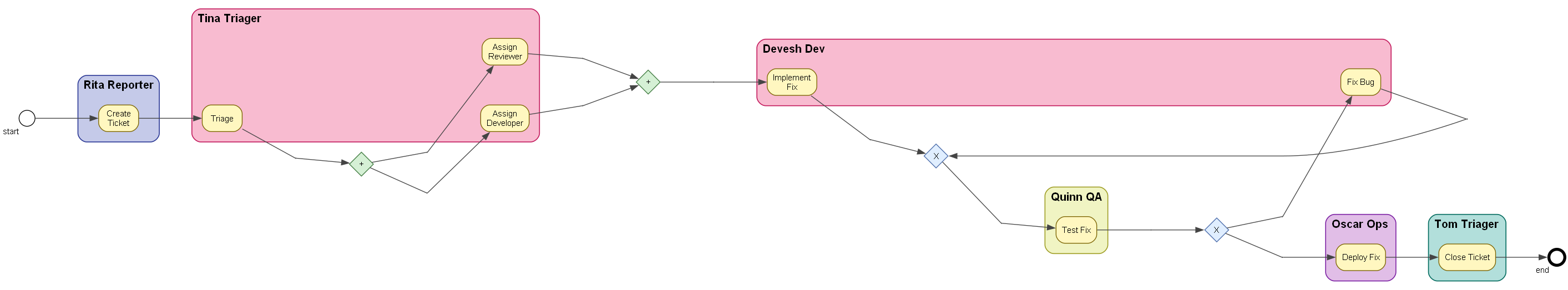}
        \caption{UCM view, with individual resources as component}
        \label{fig:initial-bpmn-resources}
    \end{subfigure}

    \caption{Models mined from the issue tracking event log.}
    \label{fig:initial}
\end{figure*}

\begin{figure*}
    \centering
    \includegraphics[width=\textwidth]{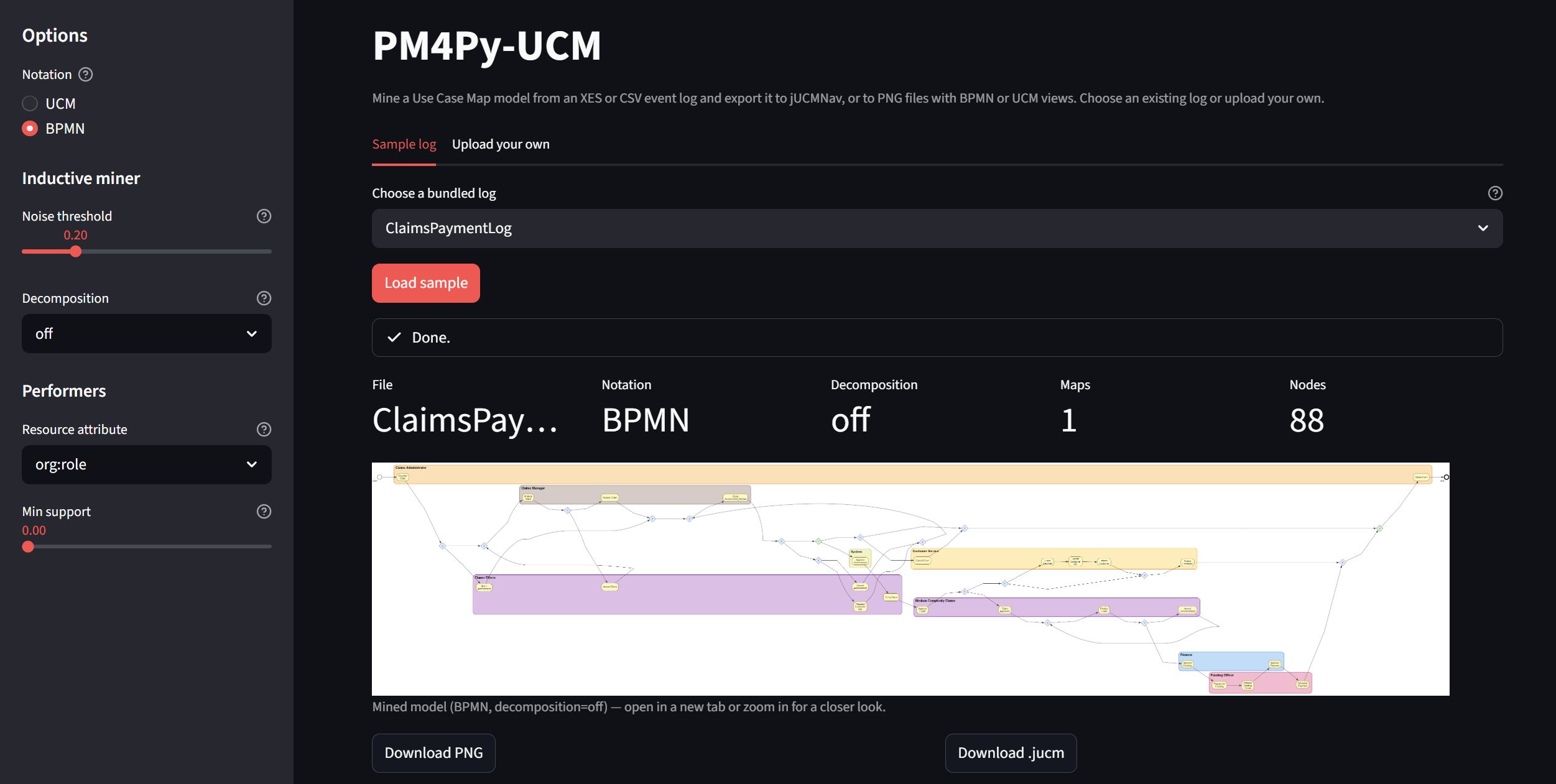}
    \caption{Claims payment: UCM view without decomposition (from within the Web interface).}
    \label{fig:Web}
\end{figure*}

\begin{figure*}
    \centering
    \includegraphics[width=1\textwidth]{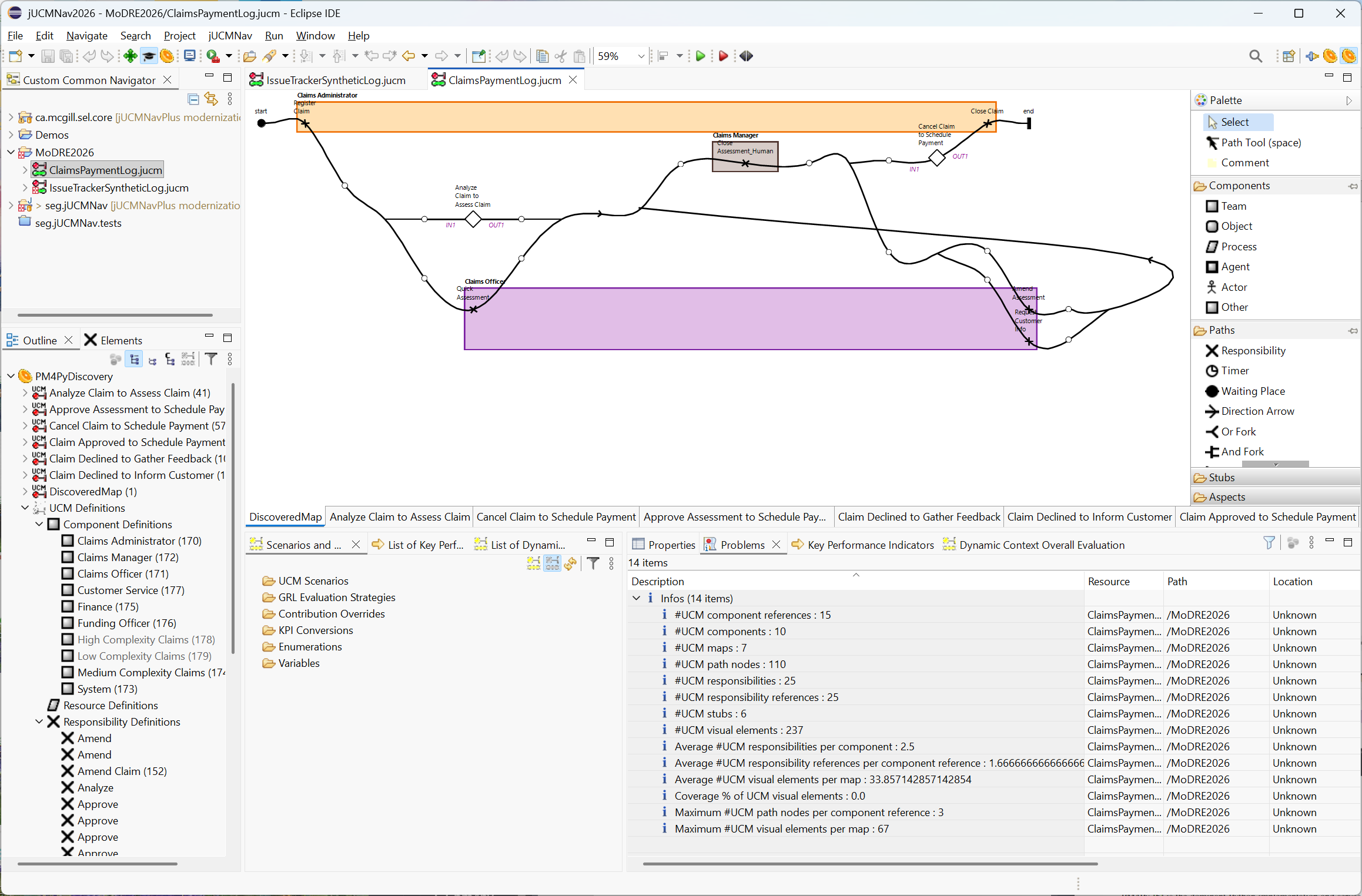}
    \caption{Claims payment: UCM model exported by \PUCM{} to a \texttt{.jucm} file and opened in \jUCMNav ~(\url{https://github.com/JUCMNAV/jUCMNavPlus}), with model metrics computed. Note that all 7 maps are there.}
    \label{fig:jUCMNav}
\end{figure*}

\begin{figure*}
    \centering
    \includegraphics[width=\textwidth]{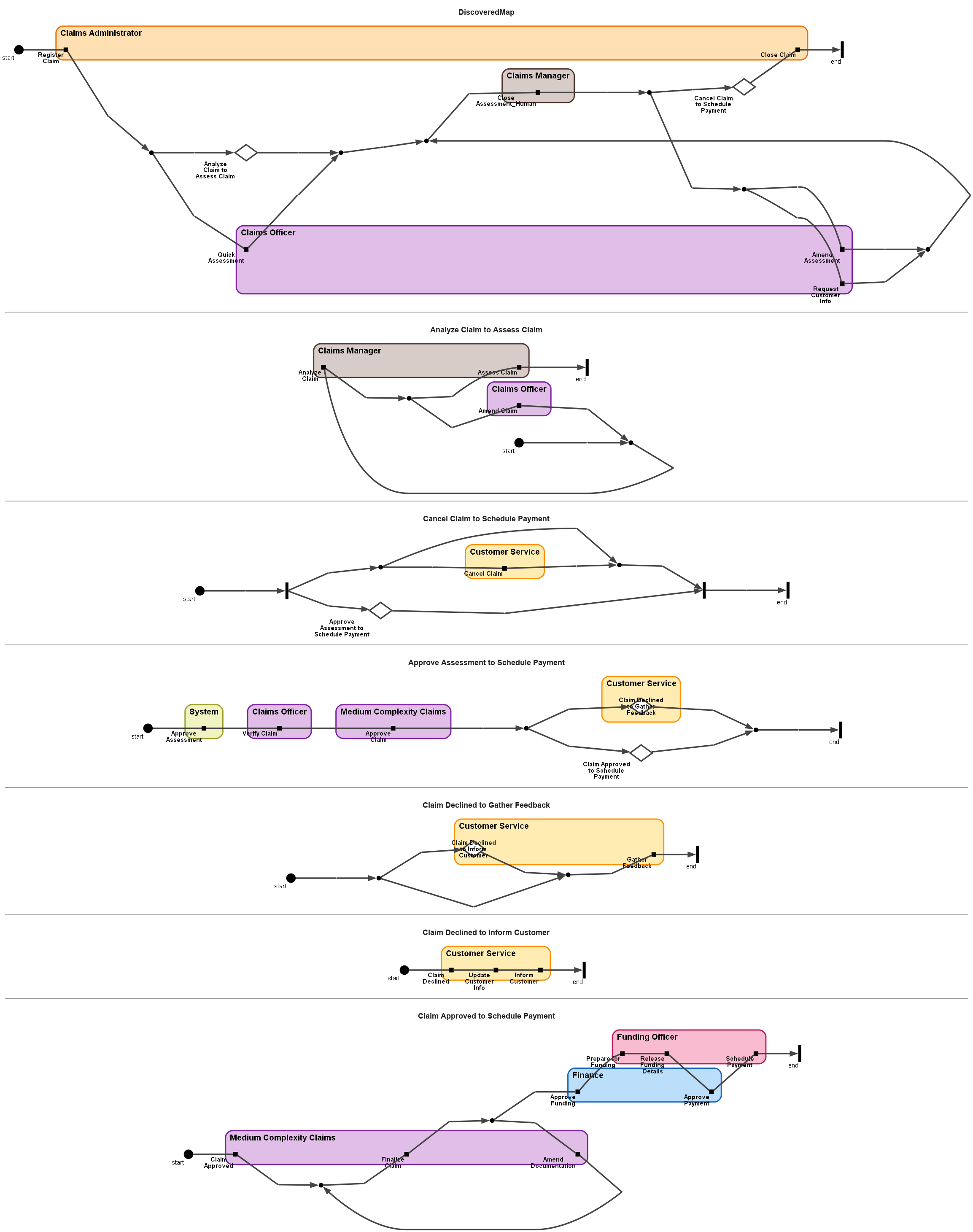}
    \caption{Claims payment: UCM view with decomposition.}
    \label{fig:decomposition}
\end{figure*}

\subsection{Initial Elicitation via Process Discovery}
The simplest \PUCM{} process model that can be extracted from an event log \One{} is one without decomposition and without components. Figure~\ref{fig:initial-ucm-nocomp} shows such a model generated from the issue tracking event log (using inductive mining \Two{} and our UCM discovery procedure \Four), visualized as a UCM \Seven. It clearly shows 9 activities (shown here with small squares on a path), and sequence, concurrency, and loop constructs that correspond to the log's process tree \Three:\\
{\small\texttt{(Create Ticket, Triage, $\land$(Assign Reviewer, Assign Developer), Implement Fix, $\circlearrowleft$(Test Fix, Fix Bug), Deploy Fix, Close Ticket)
}}

\subsection{Performer Mappings}
\label{sec:performers}
Performers can be extracted from the log (using the \textit{mode} option, see Table~\ref{tab:decompositionModes}), and injected to the UCM model \Five. Figure~\ref{fig:initial-ucm-roles} shows an example where roles were used as data source for the performer mapping. Each of the 9 activities is now bound to a UCM component. Some components (roles here) are responsible for multiple activities. Such a view definitely lacks from PM tools as it shows on the same model the \textit{who} aspect of ``who does what, and when''. RE practitioners are interested in the systems and users performing activities, not just the sequencing of the latter.

Figure~\ref{fig:initial-bpmn-resources} shows another example, using the hybrid BPMN-with-components visualization discussed earlier \Eight. In addition, resources (e.g., Tina Triager) have been used as a source of performers. This binding decision leads to differences with Figure~\ref{fig:initial-ucm-roles}. For example, although Tina Triager and Tom Triager are both members of the Triage Team, evidence shows they specialize in different activities (and might not be fully interchangeable). If further elicitation about these activities is necessary, then it would be important to have them both involved in discussions. 

\subsection{Hierarchical Decomposition}
\label{sec:hierarchy}
The variety of decomposition options discussed in Section~\ref{sec:architecture} suggests that no decomposition is canonical, and that a variety of views might be required by RE practitioners to enable effective understanding and communication. This will be illustrated using the claims payment log. Figure~\ref{fig:Web} shows the UCM model (and the Streamlit-based Web interface for \PUCM) generated, with role-based performer mapping, but without decomposition. This model contains 88 nodes and is too complex to visualize or understand as is. 

Using the decomposition \Five{} in mode auto (Table~\ref{tab:decompositionModes}) splits this model into 5 maps, but with a root map that remains overly complex. However, by providing more aggressive settings regarding the number of elements per map (\textsf{min\_leaves\_to\_decompose=3} and \textsf{max\_leaves\_per\_map=7}), the number of maps becomes higher (7) but the root map becomes much easier to understand. The generated PNG file, which names and stacks all the plug-in maps (sub-processes for their respective stubs), is illustrated in Figure~\ref{fig:decomposition}. The UCM stubs (represented as white diamonds) are the containers for sub-maps. Note also how the component colors remain consistent across the processes.

\subsection{Round-Trip to \jUCMNav{} for URN-Level Analysis}
\label{sec:roundtrip}
Mined UCM models can be exported to \texttt{.jucm} files, compliant with \jUCMNav's file format \Six. For example, Figure~\ref{fig:jUCMNav} shows \jUCMNav{} with the root map of the claims payment model with decomposition from the previous section. This is a fully editable model, that can be extended by a requirements engineer to trace its process elements to GRL goal model elements and assess alignment and consistency through jUCMNav's OCL rules~\cite{Akhigbe2016-Consistency-OCL}. Different metrics can also be computed (again through predefined OCL queries), as shown on the bottom-right of the figure. Such an import also opens the door to scenario-based testing, test generation~\cite{Kesserwan2019-UCM-testing}, and other URN-oriented capabilities briefly discussed in Section~\ref{sec:UCM}. 

Modifications to the model can also be re-imported to the UCM object model in Python, e.g., to explore other performer mappings or to visualize the results as BPMN diagrams (e.g., should stakeholders be more familiar with BPMN than UCM).

\section{Related Work}
\label{sec:related}
Different categories of related work are worth exploring here.

\subsection{Process Mining for Requirements Engineering} 
The intersection of process mining and RE has attracted growing attention, with papers arguing that mined evidence ought to feed back into requirements artifacts~\cite{Ghasemi2018-RE-PM}. In particular, Ghasemi surveyed the literature looking at the connections between process mining, goal modeling, and goal/intention mining~\cite{Ghasemi2020-REJ}, while questioning the rationality and interpretability of mined models for RE activities. This worked led to a goal-driven log filtering approach, so PM models could focus on the cases that meet certain indicator-based objectives~\cite{Ghasemi2025-GoPM}. As this applies to event logs manipulation prior to PM, this is complementary to \PUCM.

Existing PM pipelines typically discover BPMN, DFG, or Petri-net models~\cite{Berti2023-pm4py} and rely on a subsequent, often manual, translation step to obtain RE artifacts from these models. Aysolmaz et al.~\cite{Aysolmaz2018} however proposed a semi-automated approach for the generation of natural-language requirements documents from process models (mined or not), which could again complement the output of \PUCM. 

Our contribution differs in that the mined process is itself represented in a standardized RE notation (UCM/URN), which opens the door to a vast amount of other modeling and analysis activities unavailable for the other notations~\cite{Amyot2022-EMISAJ}.

\subsection{Hierarchical and Decomposed Process Discovery} 
The inductive-miner family of algorithms~\cite{LFvdA13a,LFvdA13b} produces block-structured process trees that admit a hierarchical reading, and trace-clustering approaches partition logs into behavioral variants amenable to separate discovery~\cite{SGvdA08,Trabelsi2025clustering-PM}. \PUCM{} differs in two respects here: decomposition strategies are selected along \emph{RE-meaningful} axes (process constructs, understandability) rather than fitness or precision objectives, and the resulting hierarchy remains available as UCM stubs-plugins constructs in a model editor rather than as a process tree consumed only by mining tools. 

Ghawi~\cite{Ghawi2016inductive-decomposition} also proposed combining process decomposition with the inductive-miner algorithm, but this was mainly done manually while targeting Petri Net models. \PUCM{} goes beyond this work by supporting decomposition automation via the extension of a popular library, and also via  support for performer bindings.

\subsection{Organizational and Role Mining} 
Song and van der Aalst systematized organizational mining as a sub-discipline of process mining~\cite{SvdA08} and introduced social-network discovery from event logs~\cite{vdARS05}. \PUCM{} differs in that performers are projected directly \emph{into} the requirements model (as components in UCM), where they can provide requirements engineering with combined activity/performer insights (as discussed in Section~\ref{sec:hierarchy}, rather than being rendered as a separate organizational graph orthogonal to the process model. \PUCM's support for a hybrid BPMN+components visualization also enables discussing the tool's outputs with stakeholders familiar with BPMN but unfamiliar with UCM.

\subsection{UCM and URN Tooling} 
To our knowledge, no prior tool offers a log-driven entry point into the UCM modeling and analysis chain. \PUCM{} offers that entry point, with stable round-trip engineering and mined model elements represented as first-class \jUCMNav{} citizens. Prior work by Pourshahid et al.~\cite{Pourshahid2009URN-BPM} has argued for URN as a (manual) business-process notation; \PUCM{} provides the empirical, automated, log-driven counterpart to that line of research.
\section{Limitations}
\label{sec:limitations}
This first version of \PUCM{} has many limitations that can lead to future work:
\begin{enumerate}
    \item The inductive-miner algorithm is constraining process models to being well nested; this simplifies decomposition, but at the same time this may not reflect the complexity of non-well-behaved processes. Other algorithms such as split-miner~\cite{Augusto2019-SplitMiner} (unavailable in \PMPy) could likely relax that constraint while ensuring deadlock-free concurrency in the generated models, but at the cost of a more complex decomposition solution.
    
    \item One activity could be performed by many roles or resources. The proposed aggregation strategies (Table~\ref{tab:performer-strategies}) are covering extreme cases (most frequent, first, all) but better trade-offs will be needed in practice. In particular, decomposition could allow sub-process activities to be mapped to components different from their parent maps. This would however require decomposition and performer-binding to be combined (they are performed separately in the current \PUCM).
    
    \item The decomposition parameters and user modes are tightly coupled to the notion of process trees and might not be sufficient in practice. Already, with ad hoc testing on a number of event logs, the ``aggressive'' mode shows not to be aggressive enough. Better evidence-based fine-tuning of modes is needed.
     
    \item Decomposition strategies are pragmatic (in their exploitation of process tree structures) but heuristic in nature. Clustering approaches~\cite{SGvdA08,Trabelsi2025clustering-PM} could lead to other types of decompositions that would be better based on the semantics of a process rather than solely on its structure. Proper decomposition metrics could also be used to compare their effectiveness~\cite{Milani2016criteria}.
    
    \item UCM is a rich notation where some constructs (timers, failure points, dynamic stubs, component types, etc.) are not distinguished by the current approach. Investigating the feasibility and usefulness of inferring them from event logs is left to future work.

    \item Boolean conditions on OR-forks in the UCM model are not inferred from the current PM approach, which limits the executability of mined models, e.g., for scenario execution. Decision mining~\cite{Elhami2026-DM-SLR, Rozinat2006-DecisionMining} could potentially supplement the current approach to extract such conditions. 

    \item Over 150 unit and regression test cases were developed for \PUCM. However, the library and its Web interface have not yet been used by others. There is a need for proper usability testing and empirical validation, both of the API and its parameters and of the Web interface. Limitations in practical settings, especially around adoption, need to be properly identified and addressed.

\end{enumerate}

\section{Conclusion}
\label{sec:conclusion}

This paper proposed \PUCM, and extension of the popular \PMPy{} library as a means to support the Use Case Maps notation as an output of a common process mining algorithm, as well as enable the data-driven and automated creation of UCM models in a requirements engineering context. Four important contributions were defined in Section~\ref{sec:architecture} and illustrated in Section~\ref{sec:capabilities}:

\begin{itemize}
  \item \textbf{C1.} An event-log-to-UCM discovery pipeline (Figure~\ref{fig:pipeline}). 
  \item \textbf{C2.} Configurable decomposition strategies, producing hierarchical UCM models that can be visualized (using the UCM and BPMN syntaxes) and exported as PNG bitmaps (Figure~\ref{fig:decomposition}).
  \item \textbf{C3.} Configurable performer-aware bindings of activities to components  (Figure~\ref{fig:initial}), with performers extracted from logs or provided manually.
  \item \textbf{C4.} A bi-directional importer/exporter that supports the \jUCMNav{} file format (\texttt{.jucm}), enabling round-trip engineering (Figure~\ref{fig:jUCMNav}).
\end{itemize}

In addition to the library itself, the GitHub repository also contains documentation, a Jupyter Notebook tutorial on the tool's API, and a deployed Web-based interface supporting common usages of the new library (Figure~\ref{fig:Web}, \url{https://pm4py-ucm.streamlit.app/}).

The many limitations identified in Section~\ref{sec:limitations} can lead to useful research areas focusing on a better exploitation of process mining in a requirements engineering context.

\section*{Acknowledgments}
This work is supported by an NSERC Discovery grant titled ``Requirements-Oriented Process Mining''. The author acknowledges the use of Claude Code (by Anthropic), with its Opus 4.7 model, to assist in developing, debugging, and deploying the code used in \PUCM. 

\bibliographystyle{IEEEtranDOI}
\bibliography{refs}

\end{document}